\documentclass[12pt]{article}
\usepackage[english]{babel}
\usepackage{amsmath}
\usepackage{color}
\Roman{section}
\def\RR{I\hspace{-1mm}R}
\newcommand{\eq}[1]{\begin{equation}#1\end{equation}}
\newcommand{\naw}[1]{\left(#1\right)}
\newcommand{\ket}[1]{\left|#1\right>}
\newcommand{\bra}[1]{\left<#1\right|}
\newcommand{\av}[1]{\left<#1\right>}

\newcommand{\modu}[1]{\left|#1\right|}
\newcommand{\poisson}[1]{\left\{#1\right\}}

\begin{document}

\begin{center}
\textsc{\Large{Some properties of maximally entangled ELW game}}

\emph{Katarzyna Bolonek-Laso\'n\footnote{Department of Statistical Methods, Faculty of Economics and Sociology, University of Lodz, Poland.}},
\emph{Piotr Kosi\'nski\footnote{Department of Theoretical Physics and Computer Science, Faculty of Physics and Informatics University of Lodz, Poland}}
\end{center}

\begin{abstract}
The Eisert et al. \cite{EisertWL} maximally entangled quantum game is studied within the framework of (elementary) group theory. It is shown that the game can be described in terms of real Hilbert space of states. It is also shown that the crucial properties of the maximally entangled case, like quaternionic structure and the existence, to any given strategy, the corresponding counterstrategy, result from the existence of large stability subgroup of initial state of the game.
\end{abstract}

\section{Introduction}
Quantum game theory has attracted much attention in recent decade. The most studied games are probably those introduced in Refs. \cite{EisertWL}, \cite{EisertW}, \cite{Shimamura}. 
In particular, it seems that the Eisert et al. \cite{EisertWL} version of quantum game became the paradigm for many (if not most) constructions in the domain of quantum games theory.\\ 
\indent Let us remind the basic elements of Eisert-Wilkens-Lewenstein game. One starts with classical two-players game. Each of the players must independently decide whether he/she chooses the strategy D (defect) or C (cooperate). Depending on their decision each player receives a certain payoff as described in the table below: \\
\begin{center} 
\begin{tabular}{|c|c|c|c|}\hline
\multicolumn{2}{|c|}{Strategies} & \multicolumn{2}{|c|}{Payoffs}\\
\cline{1-4}
player A & player B  & player A & player B\\
\cline{1-4}
C & C & r & r\\
C & D & s & t\\
D & C & t & s\\
D & D & p & p\\ \hline
\end{tabular}\\
\end{center}
Assuming that the payoffs obey $t>r>p>s$ (in the original paper $t=5$, $r=3$, $p=1$, $s=0$) one deals with the so-called Prisoners' Dilemma. Namely, the strategy $(D,D)$ provides the Nash equilibrium (i.e. both players conclude that he/she could not have done better by unilaterally changing his/her own strategy) whereas each of the players is doing worse than if they would both decide to cooperate.\\
\indent The quantization of the game begins by assigning the possible outcomes of the classical strategies D nad C to the basis vectors $\ket{-}$ and $\ket{+}$ in twodimensional Hilbert space. The state of the game is described by a vector in the tensor product space spanned by $\ket{++}$, $\ket{+-}$, $\ket{-+}$ and $\ket{--}$ which correspond to all possible choices of both players strategies. The initial state of the game is given by
\eq{\Psi_0=\hat{J}\ket{++}}
where $\hat{J}$ is a unitary operator known to both players; $\hat{J}$ represents entaglement and plays an important role in what follows. It is symmetric with respect to the interchange of the players.\\
\indent Strategic moves of the players are associated with unitary operators $\hat{U}_1$, $\hat{U}_2$ operating on their own qubits. The final state of the game is given by 
\eq{\ket{\Psi_f}=\hat{J}^+\naw{\hat{U}_A\otimes\hat{U}_B}\ket{\Psi_0}=\hat{J}^+\naw{\hat{U}_A\otimes\hat{U}_B}\hat{J}\ket{++}}
and the expected payoffs are computed according to
\eq{\begin{split}
& S_A=rP_{++}+pP_{--}+tP_{-+}+sP_{+-}\\
&  S_B=rP_{++}+pP_{--}+sP_{-+}+tP_{+-}
\end{split}} 
with $P_{\varepsilon\varepsilon'}\equiv\modu{\av{\varepsilon \varepsilon'|\Psi_f }}^2\equiv\modu{A_{\varepsilon \varepsilon'}}^2$.\\
In order to ensure that the quantum game entails a faithful representation of its classical counterpart (including the mixed strategies of the latter) one selects the two-bit gate operator $\hat{J}$ in the form
\eq{\hat{J}\naw{\gamma}=exp\naw{i\frac{\gamma}{2}\hat{D}\otimes\hat{D}}}
where $\hat{D}=i\sigma_2$ and $\sigma_2$ is the second Pauli matrix.\\
\indent It appears that some important properties of ELW games (for example, the resolution of Prisoner's Dilemma) depend on the choice of the manifold of admissible strategies \cite{EisertW}, \cite{Benjamin}, \cite{Du1}, \cite{Du2}, \cite{Du3}, \cite{Flitney}. The most natural choice seems to be the whole $SU(2)$ group manifold (although it may be a matter of some dispute). Assuming this is the case one can use the methods of group theory to get some insight into the structure of the game.\\
\indent The properties of the game introduced by Eisert et. al. \cite{EisertWL} depend also on the parameter $\gamma$ which determines the degree of entaglement. Particularly interesting is the case of maximal entaglement, $\gamma=\frac{\pi}{2}$. Quite recently Landsburg has shown in the series of papers \cite{Landsburg1}, \cite{Landsburg2}, \cite{Landsburg3} that the mathematical structure of maximally entangled ELW game can be described in terms of quaternionic algebra. He used this result to find the mixed-strategy Nash equilibria.\\
\indent In the present paper we describe some mathematical properties of maximally entangled ELW game using the techniques (actually, rather elementary) of group theory. 
It is well-known \cite{Birkhoff}, \cite{Jauch}, \cite{Vara} that, in order to reflect the properties of real world the quantum mechanics must be based on complex Hilbert space. In fact, if one starts with real space of states it appears that the additional structures must be introduced which makes the formalism essentially equivalent to the one based on complex space of states \cite{Stueck1}, \cite{Stueck2}, \cite{Mackey}. It is, therefore, slightly surprising that, in the case of maximal entanglement, the EWL game can be formulated entirely in terms of real Hilbert space of states. In Sec. II we give a very simple argument in favour of this conlusion.\\
\indent Further, in Sec. III we show that the existence of the so-called counterstrategies \cite{EisertW}, \cite{Benjamin} results from the fact that the stability subgroup of initially entangled state is $SO(3)$ (or, in other words, basically diagonal subgroup of $SU(2)\times SU(2)$).
In fact, the exceptional properties of maximally entangled case depend crucially on the existence of this large (as compared with $SO(2)$ for $\gamma<\frac{\pi}{2}$) stability subgroup. The sixdimensional manifold of strategies of both players reduces to threedimensional one. This makes a large qualitative difference and shows that the group theoretical structure of the game has important implications. Moreover, it seems that some of its properties (like nonexistence of pure Nash equilibria) proliferate to some neighbourhood (of the size depending on actual payoff values) of the point $\gamma=\frac{\pi}{2}$ \cite{Du1}, \cite{Du2}, \cite{Du3}.\\
\indent Finally, in Sec. IV the quaternionic structure is very shortly discussed from the group theory point of view.
It is explained that the quaternionic structure results from the pseudoreality of the basic representation of $SU(2)$ and, again, from the form of stability subgroup.

\section{The real structure of maximally entangled game}

\qquad It is well known that the complex finitedimensional irreducible group representations can be classified into three types \cite{Serre}, \cite{Fulton}. First, there are real representations which, by a suitable choice of basis, can be put in explicitly real form; second, there are pseudoreal ones which are equivalent to their complex conjugates but cannot be represented by real matrices; finally, complex ones which are inequivalent to their complex conjugates. The first two types are characterized by the existence of antiunitary operator $\mathcal{C}$ commuting with all elements of representation and obeying either $\mathcal{C}^2=I$ (real case) or $\mathcal{C}^2=-I$ (pseudoreal case) \cite{Fulton}, \cite{Giller}.\\
\indent In the case of $SU(2)$ group there is only one irreducible representation of a given dimension so all irreps are either real (integer spin) or pseudoreal (halfinteger spin). In particular, the defining spin $\frac{1}{2}$ representation is pseudoreal. In fact, if $U$ is a $SU(2)$ matrix then
\eq{U\mathcal{C}\naw{\eta}=\mathcal{C}\naw{\eta}U}
where
\eq{\mathcal{C}\naw{\eta}=\eta\sigma_2\mathcal{K};}
here $\sigma_2$ is the second Pauli matrix and $\mathcal{K}$ denotes complex conjugation (with respect to natural basis) while $\eta$ is a phase factor ($\modu{\eta}=1$); obviously $\naw{\mathcal{C}\naw{\eta}}^2=-I$.\\
\indent Consider now the $SU(2)\times SU(2)$ group. Taking for each factor the defining representation one obtains the real representation. In fact, one has
\eq{\naw{U\otimes V}\naw{\mathcal{C}\naw{\eta}\otimes\mathcal{C}\naw{\eta '}}=\naw{\mathcal{C}\naw{\eta}\otimes\mathcal{C}\naw{\eta'}}\naw{U\otimes V}}
and
\eq{\naw{\mathcal{C}\naw{\eta}\otimes\mathcal{C}\naw{\eta '}}^2=I.}
\indent Given real group representation acting in the complex Hilbert space $\mathcal{H}$ its real counterpart is constructed by applying the operator $\frac{1}{2}\naw{I+\mathcal{C}}$,
\eq{\mathcal{H}_R\equiv\poisson{\frac{1}{2}\naw{I+\mathcal{C}}\Psi \quad|\quad \Psi\in\mathcal{H}}.}
The operators representing group elements act in $\mathcal{H}_R$ as real matrices.\\
\noindent Let us now come back to our quantum game \cite{EisertWL}. The gate operator \cite{EisertWL}
\eq{J\naw{\gamma}=e^{i\frac{\gamma}{2}D\otimes D},\qquad D=i\sigma_2}
for $\gamma=\frac{\pi}{2}$ takes the form
\eq{J\naw{\frac{\pi}{2}}=\sqrt{2}\naw{\frac{I+i\sigma_2\otimes\naw{-\sigma_2}}{2}}.}
The basic vectors
\eq{\ket{++}=\left ( \begin{array}{c}
1 \\ 0\\ 0\\ 0 \end{array}\right ),\quad \ket{-+}=\left ( \begin{array}{c}
0 \\ 1\\ 0\\ 0 \end{array}\right ),\quad \ket{+-}=\left ( \begin{array}{c}
0 \\ 0\\ 1\\ 0 \end{array}\right ), \quad \ket{--}=\left ( \begin{array}{c}
0 \\ 0\\ 0\\ 1 \end{array}\right )\label{aa}}
are real (in writing out explicitly the tensor products of matrices we adopt here the convention that the first factor is inserted into the second one $\naw{a,b}^T\otimes\naw{c,d}^T=\naw{ac,bc,ad,bd}^T$). Therefore, one can write
\eq{J\naw{\frac{\pi}{2}}\ket{\varepsilon, \varepsilon '}=\sqrt{2}\naw{\frac{I+\mathcal{C}\naw{i}\otimes\mathcal{C}\naw{-1}}{2}}\ket{\varepsilon,\varepsilon '},\quad \varepsilon,\varepsilon '=\pm .}
According to the reasoning presented in this section we conclude that the amplitudes 
\eq{A_{\varepsilon\varepsilon '}=\bra{\varepsilon,\varepsilon '}J^+\naw{\frac{\pi}{2}}\naw{U_A\otimes U_B}J\naw{\frac{\pi}{2}}\ket{++}\label{a}}
are all real. Consequently, the game can be entirely defined within the framework of real space of states.

\section{The $SO(4)$ structure of maximally entangled game}
\qquad The $SU(2)\times SU(2)$ group is locally isomorphic to $SO(4)$ \cite{Fulton}. More precisely, $SO(4)$ is isomorphic to the quotient $SU(2)\times SU(2)/\poisson{\naw{I,I},\naw{-I,-I}}$. This isomorphism is given by the relation \cite{Fuji}
\eq{S=R^+\naw{U\otimes V}R\label{b}}
where
\eq{R=\frac{1}{\sqrt{2}}\left (\begin{array}{cccc}
1 & 0 & 0 & 1\\
0 & i & i & 0\\
0 & -1 & 1 & 0\\
i & 0 & 0 & -i \end{array}\right )}
and $S\in SO(4)$ is the image of $\naw{U,V}\in SU(2)\times SU(2)$ under the above isomorphism. The amplitudes (\ref{a}) take now the form
\eq{A_{\varepsilon\varepsilon '}=\bra{\varepsilon,\varepsilon '}J^+\naw{\frac{\pi}{2}}RSR^+J\naw{\frac{\pi}{2}}\ket{++}.}
Denoting
\eq{\widetilde{\ket{\varepsilon,\varepsilon '}}\equiv R^+J\naw{\frac{\pi}{2}}\ket{\varepsilon,\varepsilon '}} 
we find
\eq{
\begin{split}
& \widetilde{\ket{++}}=\frac{1}{\sqrt{2}}\naw{\frac{1+i}{\sqrt{2}}}\left (\begin{array}{c} 1 \\ 0 \\ 0 \\ 1 \end{array}\right ), \qquad \widetilde{\ket{-+}}=\frac{1}{\sqrt{2}}\naw{\frac{1+i}{\sqrt{2}}}\left (\begin{array}{c} 0 \\ 1 \\ -1 \\ 0 \end{array}\right ),\\
&\widetilde{\ket{+-}}=\frac{1}{\sqrt{2}}\naw{\frac{1+i}{\sqrt{2}}}\left (\begin{array}{c} 0 \\ 1 \\ 1 \\ 0 \end{array}\right ), \qquad \widetilde{\ket{--}}=\frac{1}{\sqrt{2}}\naw{\frac{1+i}{\sqrt{2}}}\left (\begin{array}{c} 1 \\ 0 \\ 0 \\ -1 \end{array}\right ) 
\end{split}\label{ab}}
which leads to explicitly real form of amplitudes (after cancelling the common phase factor)
\eq{A_{\varepsilon\varepsilon '}=\widetilde{\bra{\varepsilon,\varepsilon '}}S\widetilde{\ket{++}}.}
Within this framework any pair of strategies of Alice and Bob is given by a single point on $SO(4)$ manifold. Not all strategies lead, however, to different outcomes. Two element $S$, $S '$ give the same values of amplitudes if they differ by the element of stability group of $\widetilde{\ket{++}}$, $S=S'S_0$, $S_0\widetilde{\ket{++}}=\widetilde{\ket{++}}$. Therefore, the effective manifold of strategies is the coset space of $SO(4)$ divided by the stability subgroup of $\widetilde{\ket{++}}$. The latter is $SO(3)$ so the set of strategies is isomorphic to $SO(4)/SO(3)\sim S^3$.\\
\indent We see that the set of effective common strategies coincides, as a manifold, with the set of the strategies of single player. This is the reason for the existence, for any given strategy of one player, the $"$counterstrategy$"$ of the other \cite{EisertW}, \cite{Benjamin}. More precisely, the following holds true. Assume we have selected some values of the amplitudes $A_{\varepsilon\varepsilon '}$ and Alice has chosen her strategy. Then Bob can select his strategy ($"$counterstrategy$"$) in such a way that the resulting amplitudes coincide with those chosen in advance.\\
\noindent To see this let us note that the image of the stability subgroup $SO(3)$ of the vector $\widetilde{\ket{++}}$ under the isomorphism described above has the form $\naw{U,U_0UU_0^+}$, where $U\in SU(2)$ and $U_0\in SU(2)$ is some fixed element (which doesn't have to be specified more precisely). Assume $\naw{U_1,U_2}$ is a pair of strategies leading to the prescribed values of amplitudes. Finally, assume that Alice has chosen the strategy $V_1$. Note the identity
\eq{\naw{U_1,U_2}=\naw{V_1,U_2U_0U_1^+V_1U_0^+}\naw{V_1^+U_1,U_0V_1^+U_1U_0^+}.}
The last element on the right-hand side belongs to the stability subgroup of $\widetilde{\ket{++}}$. Therefore, the strategies $\naw{V_1,U_2U_0U_1^+V_1U_0^+}$ and $\naw{U_1,U_2}$ lead to the same amplitudes. So, irrespectively of the strategies chosen by Alice, Bob can always use the strategy leading to the result advantegous for him. As mentioned above the reason for this is that the stability subgroup is relatively large.\\
\indent Concluding this section let us note that the correspondence (\ref{b}) is not unique. The $R$ matrix is defined up to an orthogonal matrix. For example, writing
\eq{S'=P^+R^+\naw{U\otimes V}RP}
with    
\eq{P=\frac{1}{\sqrt{2}}\left (\begin{array}{cccc}
1 & 0 & 0 & 1\\
0 & 1 & 1 & 0\\
0 & -1 & 1 & 0\\
1 & 0 & 0 & -1 \end{array}\right )}
one finds
\eq{A_{\varepsilon\varepsilon '}={\bra{{\varepsilon,\varepsilon '}}}S'\ket{{++}},\qquad \ket{{\varepsilon,\varepsilon '}}=P^+\widetilde{\ket{\varepsilon,\varepsilon '}}\label{d}}
where $\ket{\varepsilon, \varepsilon '}$ are defined by eq. (\ref{aa})

\section{Quaternionic formulation}
\qquad The results of last section allow us to shed some light on Landsburg approach \cite{Landsburg1}, \cite{Landsburg2}, \cite{Landsburg3}. It is well known that the set of quaternions of unit norm with quaternionic multiplication is isomorphic to $SU(2)$ group. Actually, as it has been mentioned in Sec.II, all representations of $SU(2)$ are either pseudoreal (halfinteger spin) or real (integer spin). Therefore, the basic representation can indeed be rewritten in terms of quaternions of unit norm. Moreover, the quaternions are closely related to $SO(4)$ group. In fact, they are vectors in $\RR^4$. Let $r$ be an arbitrary quaternion, $q_1$ and $q_2$ - quaternions of unit norm; then
\eq{r\rightarrow q_1rq_2^{-1}\label{c}}
is the general $SO(4)$ rotation applied to r. The above formula provides an alternative proof of the relation between $SO(4)$ and $SU(2)\times SU(2)$.\\
Now, for $r=1$ eq. (\ref{c}) yields $q_1q_2^{-1}$. Fixing $q_1q_2^{-1}$ defines $q_1$ and $q_2$ up to the multiplicative factor; namely, one can make the replacement $q_1\rightarrow q_1s$, $q_2\rightarrow q_2s$, $\modu{s}=1$. Therefore, the quaternion 
\eq{q=q_1\cdot q_2^{-1}\label{s}}
is defined by the element of $SU(2)\times SU(2)$ up to the element of stability subgroup $SU(2)$. Taking into account eqs. (\ref{aa}) and (\ref{a}) we conclude that the coordinate functions of the quaternion (\ref{s}) provide the amplitudes $A_{\varepsilon\varepsilon'}$. Generally speaking, the decomposition of $SO(4)$ into $SU(2)$ factors implied by eq. (\ref{c}) may differ by an automorphism from initial decomposition in both players strategies. In fact, one can explicitly check that assuming
\eq{U_A=\left(\begin{array}{cc}
\alpha & \beta \\
-\bar{\beta} & \bar{\alpha}\end{array}\right),\qquad U_B=\left(\begin{array}{cc}
a & b \\
-\bar{b} & \bar{a}\end{array}\right)}
with $\modu{\alpha}^2+\modu{\beta}^2=1$, $\modu{a}^2+\modu{b}^2=1$, one reproduces the amplitudes (\ref{a}) via coordinate functions of $q$ provided 
\eq{\begin{split}
& q_1=Re\alpha+iIm\alpha-jRe\beta-kIm\beta\\
& q_2=Rea-iIm a+jImb+kReb.
\end{split}\label{ss}}
Standard identification of quaternion units with Pauli matrices: $i\rightarrow-i\sigma_1$, $j\rightarrow-i\sigma_2$, $k\rightarrow-i\sigma_3$ yields
\eq{\begin{split}
& U_A=Re\alpha-iIm\beta-jRe\beta-kIm\alpha\\
& U_B=Rea-iImb-jReb-kIma.
\end{split}\label{sss}}
By comparying eqs. (\ref{ss}) and (\ref{sss}) we see that $q_1$ and $q_2$ are defined by the automorphism $i\rightarrow k$, $j\rightarrow j$, $k\rightarrow -i$ and $i\rightarrow -j$, $j\rightarrow -k$, $k\rightarrow i$, respectively. 

\section{Conclusions}
\qquad We have shown that the 2-players quantum game introduced in Refs. \cite{EisertWL}, \cite{EisertW}, \cite{Shimamura} has especially interesting structure if the entaglement introduced by the gate operator takes the maximal value. The game can be then defined in the framework of real Hilbert space. What is more important, the stability subgroup of the initial state $J\naw{\frac{\pi}{2}}\ket{++}$ is relatively rich: it is $SO(3)$ or, in the $SU(2)\times SU(2)$ picture, $SU(2)$ subgroup conjugated to the diagonal one. This is the reason for the existence of $"$counterstrategy$"$, i.e. the strategy of the player which, for any given strategy of the second player, allows him/her to achieve any desired result.\\
\indent For $\gamma < \frac{\pi}{2}$ the stability subgroup is smaller. This is easily seen in the $SO(4)$ framework. The relevant counterpart of eq. (\ref{ab}) reads then
\eq{\ket{\widetilde{++}}=\frac{1}{\sqrt{2}}\left ( \begin{array}{c}
cos\naw{\frac{\gamma}{2}} \\ 0 \\ 0 \\ sin\naw{\frac{\gamma}{2}} \end{array}\right ) +\frac{i}{\sqrt{2}}\left ( \begin{array}{c} 
sin\naw{\frac{\gamma}{2}} \\ 0 \\ 0 \\ cos\naw{\frac{\gamma}{2}}\end{array}\right ).} 
The $SO(4)$ matrices are real so the stability subgroup of $\ket{\widetilde{++}}$ must leave invariant both real and imaginary parts of $\ket{\widetilde{++}}$. One easily concludes that it is the onedimensional group of rotations in 2-3 plane. In the $SU(2)\times SU(2)$ picture this subgroup consists of the matrices $e^{i\delta\sigma_3}\otimes e^{-i\delta\sigma_3}$.\\
\indent We have shown how the nice quaternionic picture of maximally entangled game, introduced by Landsburg \cite{Landsburg1}, \cite{Landsburg2}, \cite{Landsburg3} appears naturally as a consequence of group structure.\\
Finally, let us stress once more that the case of maximal entaglement is distinguished by the fact that the group of admissible strategies acts transitively on the manifold of all amplitudes; in fact, the $SO(4)$ group acts transitively on the sphere $\sum_{\varepsilon,\varepsilon '}A_{\varepsilon\varepsilon '}^2=1$. On the contrary, for $\gamma\neq\frac{\pi}{2}$ the manifold of amplitudes is sevendimensional (four complex parameters minus one normalization condition) while the effective manifold of strategies is fivedimensional (six dimensions of $SO(4)$ minus one dimension of stability subgroup).\\

\subsection*{Acknowledgement}
 We thank the anonymous referees for useful remarks and suggestions. Research of Katarzyna Bolonek-Laso\'n was supported by NCN Grant no. DEC-2012/05/D/ST2/00754.

\end{document}